\documentclass[prd,aps,showpacs,prev,preprintnumbers,floatfix,nofootinbib,twocolumn]{revtex4-1}

\pdfoutput=1

\usepackage{mathrsfs,amsmath,amsthm,amssymb,color,epstopdf,verbatim,enumerate,graphicx}
\usepackage{hyperref}
\hypersetup{colorlinks,linkcolor=cyan,citecolor=cyan,urlcolor =[rgb]{0.0,0.0,0.5}}
\usepackage[nameinlink,noabbrev]{cleveref}
\bibpunct{\color{cyan}[}{\color{cyan}]}{,}{n}{}{;}

\begin{document}

\title{Ultrahigh-energy photons from LHAASO as probes of Lorentz symmetry violations}\thanks{Phys. Rev. D 104 (2021) 063012, \url{https://link.aps.org/doi/10.1103/PhysRevD.104.063012}}
\author{Chengyi Li$\,{}^{a}$}
\author{Bo-Qiang Ma$\,{}^{a,b,c}$}\email{mabq@pku.edu.cn}
\thanks{corresponding author.}
\affiliation{${}^a$ School of Physics and State Key Laboratory of Nuclear Physics and Technology, Peking University, Beijing 100871, China\\
		${}^b$ Center for High Energy Physics, Peking University, Beijing 100871, China\\
		${}^c$ Collaborative Innovation Center of Quantum Matter, Beijing, China}
		
%\date{\today}

\begin{abstract}
The Large High Altitude Air Shower Observatory~(LHAASO) is one of the most sensitive gamma-ray detector arrays currently operating at TeV and PeV energies. Recently the LHAASO experiment detected ultra-high-energy~(UHE; $E_{\gamma}\gtrsim 100~\mathrm{TeV}$) photon emissions up to $1.4~\mathrm{PeV}$ from twelve astrophysical gamma-ray sources. We point out that the detection of cosmic photons at such energies can constrain the photon self-decay motivated by superluminal Lorentz symmetry violation~(LV) to a higher level, thus can put strong constraints to certain LV frameworks. Meanwhile, we suggest that the current observation of the PeV-scale photon with LHAASO may provide hints to permit a subluminal type of Lorentz violation in the proximity of the Planckian regime, and may be compatible with the light speed variation at the scale of $3.6\times 10^{17}~\mathrm{GeV}$ recently suggested from gamma-ray burst~(GRB) time delays. We further propose detecting PeV photons coming from extragalactic sources with future experiments, based on LV-induced threshold anomalies of $e^{+}e^{-}$ pair-production, as a crucial test of subluminal Lorentz violation. We comment that these observations are consistent with a D-brane/string-inspired quantum-gravity framework, the space-time foam model.
\end{abstract}


%\keywords{ultrahigh-energy $\gamma$-ray, Lorentz symmetry violation, string theory model, LHAASO}

\maketitle

Cosmic photons from gamma-ray sources are considered as ideal objects for investigating the Lorentz symmetry violation~(LV)~\cite{AmelinoCamelia:1996pj,AmelinoCamelia:1997gz}, since the propagation of photons at ultrahigh energies~(UHEs) above multi-TeV can be changed in an intriguing and physically meaningful manner if the dispersion relation of the photon is modified by terms suppressed by powers of the Planck mass $\mathcal{M}_{\textrm{Pl}}\simeq10^{19}~\textrm{GeV}/c^{2}$ due to the Lorentz violation. The Large High Altitude Air Shower Observatory~(LHAASO) experiment has recently observed $\gtrsim 100~\textrm{TeV}$ $\gamma$-rays from twelve astrophysical sources~\cite{Aharonian:2021pre}, for which the collaboration reported a $\gtrsim 7\sigma$ statistical significance. The sources of such high-energy photons are unknown, but preliminarily believed to be Galactic in Ref.~\cite{Aharonian:2021pre}. The observations of these UHE $\gamma$-rays in specific directions across the sky create the unique and crucial opportunity to bound certain type of Lorentz violation and can further provide key insight into the fate of the fundamental symmetry of space-time at near-Planckian energies. In this paper we discuss several implications of the detection of PeV $\gamma$-radiations in Ref.~\cite{Aharonian:2021pre} from a theoretical perspective. We suggest that, in addition to being able to limit superluminal LV effects very strongly, the LHAASO observation may be considered as a signal for a subluminal type of Lorentz violation for photons, compatible with the light speed variation~\cite{Xu:2016zxi,Xu:2016zsa,Xu:2018ien,Zhu:2021pml} as observed from recent studies on \textit{Fermi}-LAT gamma-ray bursts~(GRBs)~\cite{Atwood:2009ez,Abdo:2009aa}.

The Large High Altitude Air Shower Observatory~(LHAASO) is a new-generation ground-based extensive-air-shower detector arrays designed for cosmic-ray and $\gamma$-ray researches with unprecedented sensitivity~\cite{Aharonian:2020oil}. LHAASO is located at 4410~m above sea level at $29^{\circ}~\textrm{N}$ near the Haizi Mountain, in Sichuan, China, covering an area of 1,300,000~$\textrm{m}^{2}$. As one of the three interconnected components of the LHAASO instrument, the kilometer square array~(KM2A) has been operating continuously and stably since the end of 2019~\cite{Aharonian:2020iou}, though the construction of the KM2A detecter is still underway. Recently, the LHAASO-KM2A half-array~\cite{Aharonian:2021pre} reported detailed measurements of more than 530 photons at energies above $100~\textrm{TeV}$ up to the highest at $1.4~\textrm{PeV}$ from 12 gamma-ray sources, with the statistical significance exceeding $7\sigma$. Among these events, the most energetic photon with energy greater than $1~\textrm{PeV}$ was detected from LHAASO~J2032+4102 sited at $(\textrm{R.A.}, \mathrm{Dec})=(308.05, 41.05)$, with a $10.5\sigma$ significance~\cite{Aharonian:2021pre}. The LHAASO observation of cosmic photons at such high energies provides the unique and crucial opportunity to search for Lorentz violation, through the precise measurement and reconstruction of these UHE events. Detailed analyses of the Lorentz violation limits that can be cast with LHAASO are beyond the scope of this work. Here we limit the discussion to a few general implications on relevant Lorentz violation issues.

The breakdown of exact Lorentz symmetry at high energies is motivated by some quantum gravity~(QG) scenarios, where the usual energy-momentum relationship for standard-model particles is modified by introducing Lorentz-violating terms suppressed by $(E/\mathcal{M}_{\textrm{Pl}})^{n}$, with the power $n$ depending heavily on the underlying LV-theories taken into consideration~\cite{Shao:2010wk}. Phenomenologically, the modified dispersion relation for photons can be parametrized by~\cite{Ellis:2005wr,Zhang:2014wpb}
\begin{equation}
\label{eq:1}
E_{\gamma}^{2}=p^{2}c^{2}\left[1-s_{n}\left(\frac{pc}{E_{\textrm{LV}, n}}\right)^{n}\right],
\end{equation}
where $(E_{\gamma},p)$ indicates the four-momentum of the photon and $E_{\textrm{LV}, n}$ stands for the energy scale at which the $n$th-order LV effect set in. The LV sign factor $s_{n}=\pm1$ refers to subluminal~($s_{n}=+1$) and superluminal~($s_{n}=-1$) cases~(i.e., $s_{n}=+1$ or $s_{n}=-1$ represents that higher-energy photons travel more slowly/quickly than the low-energy photons in vacuum). Such a modified dispersion relation~\labelcref{eq:1} can affect considerably the propagation of cosmic photons with very high energies~\cite{Shao:2010wk}. For instance, the group velocity of light, $v_{\gamma}=\partial E_{\gamma}/\partial p$, is no longer a constant $c$,\footnote{Hereafter, natural units in which $\hslash,c=1$ are adopted.} but depends on the energy of that light, leading to time delays between photons with different energies. In addition, photon decay, $\gamma\rightarrow e^{+}+e^{-}$, a generally forbidden reaction in the Lorentz-symmetric Standard Model~(SM), may be allowed to occur, then can result in a cutoff to the photon flux, and moreover, the presence of Lorentz violation can alter also the energy threshold of the pair-production reaction $\gamma+\gamma_{\textrm{b}}\rightarrow e^{+}+e^{-}$, for UHE photons through interacting with low-energy background lights~($\gamma_{\mathrm{b}}$).

Considering first the photon decay, $\gamma\rightarrow e^{+}+e^{-}$, due to superluminal Lorentz violation in~\labelcref{eq:1}, we shall indicate that the LHAASO experiments are expected to place much stronger constraints to this process than previous results. As mentioned above, photon decay into $e^{+}e^{-}$ pair might be kinematically allowed at high energies in some Lorentz-violation scenarios. Detailed analyses~(see, e.g., Ref.~\cite{Martinez-Huerta:2016odc}) lead to the fact that the decay is very fast once the photon energy is above the threshold~(e.g., $\tau_{\textrm{lifetime}}\sim 10~\textrm{ns}$ for a $10~\textrm{TeV}$ photon). Utilizing the ``threshold theorem''~\cite{Mattingly:2002ba}, the threshold to valid photon self-decay for any order $n$ can be obtained as\footnote{For simplicity, we shall neglect any Lorentz violation effect in the dispersion relation of electrons, since it has been severely constrained by the absence of the vacuum Cherenkov emission.}
\begin{equation}
\label{eq:2}
\frac{4m_{e}^{2}}{E_{\gamma}^{n}\left(E_{\gamma}^{2}-4m_{e}^{2}\right)}\leqslant\frac{-s_{n}}{E_{\textrm{LV}, n}^{n}},
\end{equation}
where $m_{e}$ stands for the electron rest mass. Since the current analyses of high-energy photons from GRBs~\cite{AmelinoCamelia:2009pg,Shao:2009bv} and active galactic nuclei~(AGNs)~\cite{Li:2020uef} largely favor linearly energy-dependent LV corrections, also for simplicity, we would like to fix $n=1$ in most of the following discussions.\footnote{In the case of $n=1$ the redundant subscripts $n$ will be omitted.} In this case, taking $m_{e}\simeq511~\textrm{keV}\ll E_{\gamma}$ into account, Eq.~\labelcref{eq:2} then turns into
\begin{equation}
\label{eq:3}
p\geqslant\left(-\frac{4m_{e}^{2}E_{\textrm{LV}}}{s}\right)^{1/3},
\end{equation}
where, to first order, $E_{\gamma}\simeq p$. We can see clearly that $E_{\textrm{LV}}\rightarrow\infty$ leads to $p\rightarrow+\infty$, which is the case in the SM where the basic vertex $\gamma\rightarrow e^{+}+e^{-}$ is strictly forbidden to occur. For $s=+1$~(i.e., subluminal dispersion), the right hand side of Eq.~\labelcref{eq:3} becomes imaginary, therefore, again, no photon decay is kinematically allowed. However, for $s=-1$, Eq.~\labelcref{eq:3} implies that any high-energy photons with their momenta greater than $p_{(\textrm{th})}=(4m_{e}^{2}E_{\textrm{LV}})^{1/3}$ would rapidly decay to $e^{+}e^{-}$ pair~\cite{Jacobson:2002hd}, resulting in a significant decrease of the photon flux at UHEs beyond which no photons should reach the Earth. This leads to a sharp cutoff in the observed spectra of $\gamma$-rays and as a result, we are not supposed to detect photons with energy higher than $E_{\gamma (\textrm{th})}\simeq (4m_{e}^{2}E_{\textrm{LV}})^{1/3}$ from distant astronomical sources. Therefore, any observations of UHE cosmic photon events can set a lower limit on the energy scale of possible \textit{superluminal} Lorentz violation. Specifically, from Eq.~\labelcref{eq:3} we have
\begin{equation}
\label{eq:4}
E_{\textrm{LV}}^{(\textrm{sup})}\gtrsim9.57\times 10^{32}~\textrm{eV}\left(\frac{E_{\gamma}}{\textrm{PeV}}\right)^{3}.
\end{equation}
Actually, utilizing the above scenario, the observations of the $50~\textrm{TeV}$ and $80~\textrm{TeV}$ photons from the Crab Nebula~\cite{Martinez-Huerta:2016azo} have already cast strong constraints on Lorentz-violating $\gamma$-decays, with the relevant scale constrained to the level $E_{\textrm{LV}}^{(\textrm{sup})}\gtrsim 10^{20}~\textrm{GeV}$. A stronger result comes from recent observations of gamma-rays around $100~\mathrm{TeV}$ with the High Altitude Water Cherenkov~(HAWC) Observatory~\cite{Albert:2019nnn}. There a $2\sigma$ constraint is obtained to be $E_{\textrm{LV}}^{(\textrm{sup})}>2.22\times 10^{22}~\textrm{GeV}$, over 1800 times the Planck energy. However, compared to these previously observed multi-TeV events, the UHE $\gamma$-rays with energies up to $1.4~\textrm{PeV}$ lately detected by LHAASO turn out to be the most high-energetic part of the currently observed electromagnetic~(EM) spectrum. So we could have an improvement of 2 to 4 orders of magnitude over previous limits to superluminal $\gamma$-decays. To see this clearly, one can insert the energy $E_{\gamma(\max)}=1.42~\textrm{PeV}$ of the highest-energy event from LHAASO~J2032+4102~\cite{Aharonian:2021pre} into Eq.~\labelcref{eq:4} to make an estimation of
\begin{equation}
\label{eq:5}
E_{\textrm{LV}}^{(\textrm{sup})}\gtrsim2.74\times 10^{24}~\textrm{GeV},
\end{equation}
over 220,000 times the Planck energy scale. If we take into account the uncertainty of $E_{\gamma(\max)}$, the lower limit to $E_{\textrm{LV}}^{(\textrm{sup})}$ turns into $2.74^{+0.82}_{-0.69}\times 10^{24}~\textrm{GeV}$. In fact we can also see from Eq.~\labelcref{eq:2} and/or Eq.~\labelcref{eq:3} that the lower bounds on $E_{\textrm{LV}}^{(\textrm{sup})}$ become tighter with the increase in the observed photon energy $E_{\gamma}$ by a factor of $E_{\gamma}^{1+2/n}$~(e.g., $E_{\gamma}^{3}$ in the case of $n=1$), so it is natural to expect that the LHAASO PeV photon can lead to a more stringent constraint to superluminal dominant effects.\footnote{Shortly after the current paper was made public, the LHAASO collaboration article on constraining superluminal effects was appeared~\cite{Cao:2021pre}, which primely complements the work here presented.} We highlight that the Lorentz violation bound given in Eq.~\labelcref{eq:5} is likely to be the strongest constraint one could attain from the LHAASO data. Here, we further point out that the LHAASO experiment can put more stringent limits to certain Lorentz-violation theories. For instance, considering the framework of the standard-model extension~(SME)~\cite{Colladay:1996iz,Colladay:1998fq} with dimension-5~(\textit{CPT}-odd) operators~\cite{Myers:2003fd}, an effective field-theoretic~(EFT) treatment of Lorentz violation, where the dispersion relation of the photon field is nothing but an analogy to the generalized relation~\labelcref{eq:1} by the replacement
\begin{equation}
\label{eq:6}
\frac{s}{E_{\textrm{LV}}}\mapsto\mp\frac{2\xi}{\mathcal{M}_{\textrm{Pl}}},
\end{equation}
the LHAASO PeV photons can be used to evaluate the related limits to the LV coefficient $\xi$ entailed in the SME framework. The resulting constraint on $\xi$ turns out to be $\xi\lesssim2.23\times 10^{-6}$, on account of Eq.~\labelcref{eq:5}, far below the natural order~($\xi=\mathcal{O}(1)$) as one expects. In addition, the SME with dispersion relations~\labelcref{eq:6} predicts also the birefringence effect~\cite{Carroll:1989vb} which has been strongly constrained by experiments~\cite{Kostelecky:2008bfz}, over 11 orders of magnitude~\cite{Wei:2019nhm} than the $\gamma$-decay constraint derived here.

As an aside, related bounds for quadratically energy-dependent LV terms in~\labelcref{eq:1} from the LHAASO 1.42~PeV event can be derived also. By comparing the contributions to the generalized modified dispersion relation from the linear and quadratic LV corrections, we immediately find that the constraint~\labelcref{eq:5} can be translated into
\begin{equation}
\label{eq:7}
E_{\textrm{LV},n=2}^{(\textrm{sup})}\gtrsim\left(E_{\gamma}E_{\textrm{LV},n=1}^{(\textrm{sup})}\right)^{1/2}\approx 1.97\times 10^{15}~\textrm{GeV},
\end{equation}
where $E_{\textrm{LV},n=1}^{(\textrm{sup})}\equiv E_{\textrm{LV}}^{(\textrm{sup})}$ is given by Eq.~\labelcref{eq:5}. Such result is compatible with that obtained later by the LHAASO collaboration in Ref.~\cite{Cao:2021pre}, to which we point the reader for a wider discussion of this type of bounds. In this quadratic~($n=2$) case, constraints on the \textit{CPT}-even coefficients in the SME framework can be got.

Another interesting implication of LHAASO measurements we would like to discuss comes from the consideration of pair-production, $\gamma+\gamma_{\textrm{b}}\rightarrow e^{+}+e^{-}$, for an energetic gamma-ray ($\gamma$) through interacting with a soft background photon field~($\gamma_{\textrm{b}}$). In cosmological cases, the universal background EM-radiation field could be the cosmic microwave background~(CMB) radiation or the extragalactic background light~(EBL). The classical prediction of such reactions in special relativity~(SR) is that there is a threshold energy of low-energy lights, to absorb a high-energy $\gamma$-ray photon~(of energy $E_{\gamma}$), of the form\footnote{In special relativity, the threshold~\cite{AmelinoCamelia:2000zs} for the head-on collision between a soft photon of energy $\varepsilon$ and an energetic particle of energy $E$ ($E\gg\varepsilon$, and of rest mass $m_{E}$) is given by
\begin{equation*}
\varepsilon_{(\textrm{th})}\simeq\frac{(m_{1}+m_{2})^{2}-m_{E}^{2}}{4E},
\end{equation*}
where $m_{i}~(i=1, 2)$ stands for the mass of the outgoing particle. For $\gamma\gamma\rightarrow e^{+}e^{-}$, the threshold reduces to $m_{e}^{2}/E_{\gamma}$, \textit{i.e.},~\labelcref{eq:8}.}
\begin{equation}
\label{eq:8}
\varepsilon_{\textrm{b}(\textrm{th})}=\frac{m_{e}^{2}}{E_{\gamma}},
\end{equation}
where $\varepsilon_{\textrm{b}}$ denotes the energy of a soft background photon. The production of $e^{+}e^{-}$ can then take place beyond this threshold, i.e., $\varepsilon_{\textrm{b}}\geqslant\varepsilon_{\textrm{b}(\textrm{th})}$. This implies that with the energy $E_{\gamma}$ of high-energy photon increasing, the threshold~\labelcref{eq:8} to turn on the reaction decreases, resulting in an attenuation of the photon flux at UHEs. In order to make an estimation, we consider here the low-energy background lights as the near-infrared CMB~\cite{Fixsen:2009ug} photons which have been well resolved. Using the thermodynamic temperature of the CMB radiation $T_{\textrm{CMB}}\simeq2.7~\textrm{K}$, which corresponds to the (peak) energy $\varepsilon_{\textrm{CMB}}\approx6.6\times 10^{-4}~\textrm{eV}$, and without taking into account the mean free paths of energetic photons in the cosmic background, we can arrive at the energy limitation of any high-energy cosmic photons as $E_{\gamma}\approx4\times 10^{14}~\textrm{eV}$. This indicates that most UHE photons would be annihilated by those low energy ``absorbers'' on their way to the Earth, and again there should be a cutoff in the astrophysical $\gamma$-ray spectrum, and thus it is very unlikely to observe photons with energy higher than $400~\textrm{TeV}$, especially $\gtrsim 1~\textrm{PeV}$, from celestial sources if Lorentz symmetry holds. However, eight~(or ten) out of twelve LHAASO UHE sources reported in Ref.~\cite{Aharonian:2021pre}~(see Table 1 therein) exceed this limitation of the photon energy set by special relativity, and more interestingly, for all twelve sources the effect of pair-production appears to be small even at the highest energies (see Fig.~1 in Ref.~\cite{Aharonian:2021pre}). Therefore at least, in terms of the energy limitation of the photon surviving from absorptions by the CMB relic radiation, we suggest, with tentative arguments provided below, that the LHAASO detection of these PeV photons might be a signal for a \textit{subluminal} type of Lorentz violation. We notice that with Lorentz violation, the modified photon dispersion relation~(cf.,~\labelcref{eq:1}) will deform the standard kinematics in a physical meaningful manner, i.e., in such a way that, utilizing again the ``threshold theorem''~\cite{Mattingly:2002ba}, the usual pair-production threshold~\labelcref{eq:8} turns into
\begin{equation}
\label{eq:9}
\varepsilon_{\textrm{b}(\textrm{th})}^{\textrm{LV}}=\frac{m_{e}^{2}}{E_{\gamma}}+\frac{1}{4}\frac{sE_{\gamma}^{2}}{E_{\textrm{LV}}}.
\end{equation}
It is easily seen that $E_{\textrm{LV}}\rightarrow\infty$ gives $\varepsilon_{\textrm{b}(\textrm{th})}^{\textrm{LV}}\rightarrow m_{e}^{2}/E_{\gamma}$, thus the Lorentz-symmetric case in Eq.~\labelcref{eq:8} is recovered. For $s=-1$~(i.e., superluminal propagation), the $\gamma$-decay process dominates, so we shall focus on the case of $s=+1$. Inspecting Eq.~\labelcref{eq:9} we immediately find that there is a global minimum on $\varepsilon_{\textrm{b}(\textrm{th})}^{\textrm{LV}}$, which could be reached at a critical photon energy $E_{\gamma}=E_{\gamma(\textrm{cr})}$ by requiring $\partial\varepsilon_{\textrm{b}(\textrm{th})}^{\textrm{LV}}/\partial E_{\gamma}=0$. The presence of this minimal threshold can lead to peculiar new phenomena~\cite{Jacob:2008gj} that differ from that in special relativity. Specifically, for photons above the critical point
\begin{equation}
\label{eq:10}
E_{\gamma(\textrm{cr})}=\left(\frac{2m_{e}^{2}E_{\textrm{LV}}}{s}\right)^{1/3},
\end{equation}
with the increase in the hard $\gamma$-ray photon energy $E_{\gamma}$, the reaction threshold~\labelcref{eq:9} also increases. Since the number density of the low-energy background photons decreases with energy, there are less background lights to interact with the energetic photon above the critical $E_{\gamma(\textrm{cr})}$ and therefore, a ``reemergence'' of the energetic photons in the astrophysical spectra of $\gamma$-rays may be expected~\cite{Shao:2010wk} due to subluminal Lorentz violation impacts. This UHE $\gamma$-ray spectrum anomaly may serve as a way to account for the detection of PeV photons, reported by LHAASO, of energies well above the classical limit value, and thus this observation may be considered as a hint to permit a subluminal Lorentz symmetry breaking at the Planck-scale level,\footnote{Again we emphasize that this should be viewed as a suggestion, instead of being considered as a definite, rigorous or enough safe conclusion, for reasons we shall comment on later, from the aspect of the attenuation distance of the process $\gamma\gamma\rightarrow e^{+}e^{-}$.} though detailed consistency check using data is required before drawing such claim conclusively.

Here, we would like to further remark that a recently suggested Lorentz-violating picture~\cite{Xu:2016zxi,Xu:2016zsa,Xu:2018ien,Zhu:2021pml} emerging from \textit{Fermi}-LAT GRB time-delay data may be consistent with the above observation. Recall that in Refs.~\cite{Xu:2016zxi,Xu:2016zsa,Xu:2018ien,Zhu:2021pml}, analyses on GRB photons led to a suggestion of a \textit{subluminal} light speed variation in vacuum with the linear-order Lorentz violation energy scale determined to be
\begin{equation}
\label{eq:11}
E_{\textrm{LV}}^{(\textrm{sub})}\simeq3.6\times 10^{17}~\textrm{GeV}.
\end{equation}
Now by substituting this LV scale~\labelcref{eq:11} into the expression of the critical $\gamma$-ray energy~\labelcref{eq:10}, we find
\begin{equation}
\label{eq:12}
E_{\gamma(\textrm{cr})}\approx5.7~\textrm{TeV},
\end{equation}
and the corresponding soft photon threshold energy reaches its minimum $\varepsilon_{\textrm{b}(\textrm{th})}^{\textrm{LV}}\rvert_{\min}\approx6.8\times 10^{-2}~\textrm{eV}$, over 2 orders of magnitude the typical energy of the CMB radiation. This fact indicates that the CMB relic photons of which energies are in general lower than this minimal threshold can never interact with the high-energy $\gamma$-rays.\footnote{Similar viewpoints are clarified by H.~Li and one of us~\cite{Li:2021cdz}, where this phenomenon is termed as ``optical transparency''. More insightful discussions on that can also be found there.} Moreover, as mentioned before, the background EM-radiation fields could be also the faint diffusing extragalactic lights, of which the EM-spectrum varies from the radio band~($\gtrsim10^{-7}~\textrm{eV}$) to the ultraviolet band~($\lesssim10^{2}~\textrm{eV}$). So these lights may absorb the energetic $\gamma$-ray photons with energies around the critical value~\labelcref{eq:12} through pair-production. Keep in mind, however, that the threshold~\labelcref{eq:9} increases dramatically with increasing photon energy $E_{\gamma}$. Considering the most energetic event observed by LHAASO, then by inserting its energy $E_{\gamma(\max)}=1.42~\textrm{PeV}$ into Eq.~\labelcref{eq:9}, we can get $\varepsilon_{\textrm{b}(\mathrm{th})}^{\textrm{LV}}\simeq1.4~\textrm{keV}$, far above the attainable energy of EBLs. This fact promises the recovery of cosmic gamma-rays at PeV energies, making it possible for these photons to travel from celestial sources to the Earth, without being attenuated by CMB radiations or diffuse EBL fields during propagation, to be finally observed by LHAASO. We deduce, therefore, that the detection of PeV spectra of $\gamma$-ray sources at the LHAASO may be compatible with the previous suggestion of light speed variation at the scale of~\labelcref{eq:11}, proposed in Refs.~\cite{Xu:2016zxi,Xu:2016zsa,Xu:2018ien,Zhu:2021pml} from GRBs.

However, we should make it clear that the discussion above and our proposal of a subluminal Lorentz violation from threshold anomalies are based upon neglecting finite attenuation distance of the pair-production process, or technically, without taking into account the relatively large free paths of energetic photons in the cosmic background. Normally, assuming conventional Lorentz-invariant kinematics, the mean free path of a photon with energy $E_{\gamma}\sim 1~\textrm{PeV}$ is $\lambda_{\gamma}(E_{\gamma})\sim 10~\textrm{kpc}$~\cite{Berezinsky:2016feh}. Thus for sources within our Galaxy~(of its typical size $\sim 34~\textrm{kpc}$), as believed in Ref.~\cite{Aharonian:2021pre} to be the origin of these LHAASO photons, the $\gamma\gamma_{\textrm{b}}$ absorption should still be of relevance to some extent and can then lead to an attenuation of UHE photons from their corresponding sources, though one should notice that the survival probability of a $\sim 1~\textrm{PeV}$ photon emitted from a source, with enough small redshift $z\simeq 0$, for instance, at the distance $L\simeq 10~\textrm{kpc}$, is
\begin{equation}
\label{eq:13}
P_{\gamma\rightarrow\gamma}(E_{\gamma},z)=\textrm{e}^{-\tau_{\gamma}(E_{\gamma},z)}\simeq 0.37,
\end{equation}
where the optical depth $\tau_{\gamma}(E_{\gamma},z)$, which quantifies the dimming of the $\gamma$-ray spectrum of the source at redshift $z$, can be expressed, in the limit of a local Universe~($z\simeq 0$) as considered here, as follows
\begin{equation}
\label{eq:14}
\tau_{\gamma}(E_{\gamma},LH_{0})=\frac{L}{\lambda_{\gamma}(E_{\gamma})}\simeq 1,
\end{equation}
where the source distance $L=z/H_{0}$ , with $H_{0}$ the Hubble constant. More accurately, the photon survival probability~\labelcref{eq:13} has been computed to be larger than $37\%$ for any value of $E_{\gamma}$ at distances $\lesssim 8~\textrm{kpc}$~\cite{DeAngelis:2013jna}. This indicates that actually,  we still have a large probability to observe these PeV photons at the LHAASO in the absence of any effect of Lorentz violation, if these photons come from sources that are not very far away from us such that they did not suffer from the photon absorption very severely. Hence we realize that the situation should be reassessed prudently, since the observable consequences of pair-production affected by subluminal Lorentz violation in any astrophysical spectrum associate not only with the photon energy $E_{\gamma}$, that is the threshold effect~\labelcref{eq:9} as considered, but also with the redshift $z$ of the source, and in fact, at the present stage, one cannot certain about the existence of such a Planck-scale Lorentz violation with positive $s$~(subluminal type), in view of the comments above. Having said that, analyses on TeV emissions from blazars or AGNs~\cite{Horns:2016vfv} do not deny that there are signatures of anomalous deviations from the expected absorption during $\gamma$-ray propagation. In this sense, we just suggest that there exists a possibility that the latest detection of extremely high-energy photons by LHAASO implies a highly-suppressed deviation from exact Lorentz invariance. To settle firmly this issue, one should employ more elaborate techniques involving the optical depth in~\labelcref{eq:13} to describe quantitatively the possible attenuation anomalies of photon flux due to LV-deformed pair-production~(see, e.g., the strategy implemented in Ref.~\cite{Biteau:2015xpa} and elsewhere). This is, however, beyond the scope of the present work that is devoted to mentioning a few possible implications of new physics from the LHAASO discovery, so we will not involve any further discussion on searching for anomalous transparency of the cosmic background to LHAASO $\gamma$-rays due to LV impacts. As for now, to capture correctly the spirit of our suggestion, we stress again that it is at least appropriate to consider the LHAASO PeV event as presenting potential hint for Lorentz violation with subluminal nature at about $\sim 10^{17}~\textrm{GeV}$ to $10^{19}~\textrm{GeV}\sim\mathcal{M}_{\textrm{Pl}}$, the characteristic mass scale of any presumed QG.

Notice also, as explained, that the threshold deformation~\labelcref{eq:9} can lead to a reduction of the cosmic lights that could absorb the $\gamma$-rays above the critical point, thus the $\gamma$-ray photons could propagate farther in the cosmos than one usually expects in the Lorentz-invariant theory, a case in which these photons would be greatly attenuated through absorptions by soft photons. This indicates that UHE $\gamma$-rays, at PeV scale, originating in extragalactic active regions, like the center of distant galaxies, may also be able to travel across distances of hundreds of thousands of light-years and reach the Earth. Based on this hypothesis, as stated also in Ref.~\cite{Li:2021cdz}, we propose to search for PeV-range photons from extragalactic objects in the future with LHAASO, which, if these extragalactic $\gamma$-rays at such great energies are observed experimentally, would be a strong sign of subluminal Lorentz violation as suggested above. Of course, one may notice that very strong constraints have been derived from the current absence of significant signatures of $\gamma\gamma_{\textrm{b}}\rightarrow e^{+}e^{-}$ threshold shifts expected by LV~(see, for instance, Refs.~\cite{Biteau:2015xpa,Lang:2018yog}) using TeV gamma-ray observations of blazars. Especially, in Ref.~\cite{Lang:2018yog}, the strongest exclusion limit to subluminal signatures of LV is obtained to be higher than about ten times the Planck scale $\mathcal{M}_{\textrm{Pl}}$. However it has been shown~\cite{Li:2021cdz} that\footnote{Assume that the gamma-ray photon is attenuated by CMB for concreteness. Note that for $E_{\textrm{LV}}>4\times 10^{23}~\textrm{GeV}$ in the subluminal case, pair-production threshold anomalies still exist, but could have distinct behavior, see, Ref.~\cite{Li:2021cdz} for details.} as long as $E_{\textrm{LV}}^{(\textrm{sub})}<4\times 10^{23}~\textrm{GeV}$ the threshold anomalies exist in the sense that the cosmic background light is more transparent to subluminal photons with respect to the standard prediction in special relativity. Therefore even if one adopts a more stringent bound on $E_{\textrm{LV}}$, such as $E_{\textrm{LV}}^{(\textrm{sub})}\gtrsim 10^{20}~\textrm{GeV}$ provided by Ref.~\cite{Lang:2018yog}, the degree of freedom for the above condition of permitting threshold shifts is still large. So we emphasize that our discussion of the optical transparency is applicable for the subluminal LV case with $E_{\textrm{LV}}\lesssim 10^{23}~\textrm{GeV}$, which is within all phenomenological constraints, without the specific requirement of Eq.~\labelcref{eq:11} suggested in Refs.~\cite{Xu:2016zxi,Xu:2016zsa,Xu:2018ien,Zhu:2021pml}. On the other hand, gamma-ray observatories before LHAASO can only cover energies up to hundreds of TeV due to the limit of the collection area. Thus previous LV studies are usually limited by the maximal energy that can be measured by the experiments. However, with the onset of LHAASO, we may have chance to observe reductions of the cosmic $\gamma$-opacity, since LHAASO is a promising instrument of opening the unprecedented window of observations on $\gamma$-ray photons ranging from several tens of TeV up to a few PeV. There could be possibilities that some of the upcoming PeV events might come from extragalactic sources. This would update the data set of these objects known in the past, and might then provide convincing signals of subluminal LV.

It is also worth noticing, on the theoretical side, that certain models of QG could account for all or part of the observed phenomenology in astroparticle physics to date, including the latest LHAASO result. Since we have seen that LV models that admit an EFT formulation is in tension with various observations~(such as those constraints from the absence of $\gamma$-decay, as mentioned, or see, \textit{e.g.},~\cite{Shao:2010wk}), those QG theories that cannot be embedded into the EFT framework, specifically the D-brane/string model of space-time foam~\cite{Ellis:2004ay,Ellis:2008gg,Li:2009tt}, as inspired initially from the Liouville-string approach toward QG~\cite{Ellis:1992eh}, deserve our attentions. This framework, that appears as a
fully developed QG scenario that could ``effectively'' violate Lorentz invariance, stems from the type-IA string theory~\cite{Schwarz:1999xj} in particular, according to which the quantum-gravitational fluctuation in the space-time background---``space-time foam''---is modeled as a gas of point-like D0-brane~(``D-particle'') defects in the bulk space-time of a higher dimensional cosmology, in which the observable Universe is viewed as a D3-brane. The vacuum behaves like a dispersive gravitational medium that may have virtual structure, thus has a tendency to slow down the energetic photons. QG effects therefore motivate nontrivial \textit{subluminal} corrections to the standard relativistic relation of photons that depends \textit{linearly} on the photon energies~\cite{Li:2021gah,Li:2021eza}
\begin{equation}
\label{eq:15}
p^{2}-E_{\gamma}^{2}\simeq\frac{pE_{\gamma}^{2}}{\mathcal{M}_{\textrm{QG-D-foam}}},
\end{equation}
where $\mathcal{M}_{\textrm{QG-D-foam}}$ denotes the stringy quantum-gravity scale, being in general redshift-$z$-dependent for the~(non-uniform) D-particle foam. It has been noted~\cite{Li:2021gah,Li:2021eza} that this stringy QG framework is able to explain the light speed variation~\cite{Xu:2016zxi,Xu:2016zsa,Xu:2018ien,Zhu:2021pml} from GRB multi-GeV photons in a natural fashion, while being compatible with many other astrophysical results available to date, such as severe limits from $\gamma$-decays and/or birefringent effects. Detailed discussions can be found in Refs.~\cite{Li:2021gah,Li:2021eza}. Here, we want to further indicate that this D-brane foam model is also consistent with the LHAASO experiment. First, the QG-foam modified in-vacuo propagation of light is subluminal, as seen from Eq.~\labelcref{eq:15}, without any superluminal propagation predicted, which implies that photons are \textit{stable}~(i.e., do \textit{not} decay), and therefore, it is straightforward to expect that no observation would favor $\gamma$-decay processes to take place at high energies, and hence very strict constraints to such phenomena~(and as a consequence, to superluminal type of Lorentz violation) should be imposed by experiments. This is exactly the situation for current observational results, such as Eq.~\labelcref{eq:5}, as derived, and those obtained in Ref.~\cite{Cao:2021pre}, using the latest LHAASO events. Second, there might exist threshold anomalies, inferred intuitively from~\labelcref{eq:15}, analogous to that of Eq.~\labelcref{eq:9}, for the $\gamma\gamma_{\mathrm{b}}\rightarrow e^{+}e^{-}$ pair-production annihilation in such theory. In this respect, the theoretical prediction on the cosmic propagation of UHE $\gamma$-rays might coincide well with the observation of PeV events found by the latest experiment. Hence, we conclude that the D-particle/string foam explanation of light speed variation, proposed recently in Refs.~\cite{Li:2021gah,Li:2021eza}, can be supported as well by the LHAASO discovery.

To sum up, given that the LHAASO experiment found evidence of PeV-scale $\gamma$-radiations at $\gtrsim 7\sigma$ from twelve astrophysical sources, these ultrahigh-energy photons can serve as very sensitive probes of violation of space-time Lorentz symmetry. We show that the detection of such high-energy gamma-rays can constrain the superluminal Lorentz-violating photon decay to a higher level, and further exclude hopefully those beyond-SM scenarios that predict such peculiar phenomenon to occur. As a primary study, we extract a constraint to superluminal Lorentz-violation scale from the LHAASO $1.4~\textrm{PeV}$ event to $\sim 2.7\times 10^{24}~\textrm{GeV}$, improved by 2 to 4 orders of magnitude than those in Ref.~\cite{Albert:2019nnn} and Refs.~\cite{Martinez-Huerta:2016azo,Schreck:2013paa}. Meanwhile, we suggest that the UHE $\gamma$-ray observation with LHAASO could be a signal to permit a subluminal deformation of Lorentz symmetry at microscopic~(Planckian) scales, and this may be compatible with the light speed variation with the determined scale $E_{\textrm{LV}}^{(\textrm{sub})}=3.6\times 10^{17}~\textrm{GeV}$ as observed from \textit{Fermi}-GRBs. We further comment that the string-inspired model of space-time foam, a fully developed quantum-gravity framework capable of explaining the light speed variation from GRBs, can be supported also by the LHAASO observation of PeV photons.

This work is supported by National Natural Science Foundation of China (Grant No.~12075003).

\end{document}